\documentclass[5p,twocolumn]{elsarticle}

\usepackage{hyperref}
\usepackage{graphicx}
\usepackage{amsfonts}
\usepackage{amsmath}
\usepackage[switch]{lineno}
\usepackage{multirow}

\usepackage{float}

\begin{document}

\begin{frontmatter}

\title{Transformer with Peak Suppression and Knowledge Guidance for Fine-grained Image Recognition}


\author[mymainaddress]{Xinda Liu}
\author[mymainaddress]{Lili Wang\corref{mycorrespondingauthor}}
\author[mysecondaryaddress]{Xiaoguang Han}
\cortext[mycorrespondingauthor]{Corresponding author}

\address[mymainaddress]{State Key Laboratory of Virtual Reality Technology and Systems, Beihang University, Beijing, China}
\address[mysecondaryaddress]{Shenzhen Research Institute of Big Data, Shenzhen, China}

\begin{abstract}
Fine-grained image recognition is challenging because discriminative clues are usually fragmented, whether from a single image or multiple images. Despite their significant improvements, the majority of existing methods still focus on the most discriminative parts from a single image, ignoring informative details in other regions and lacking consideration of clues from other associated images. In this paper, we analyze the difficulties of fine-grained image recognition from a new perspective and propose a transformer architecture with the peak suppression module and knowledge guidance module, which respects the diversification of discriminative features in a single image and the aggregation of discriminative clues among multiple images. Specifically, the peak suppression module first utilizes a linear projection to convert the input image into sequential tokens. It then blocks the token based on the attention response generated by the transformer encoder. This module penalizes the attention to the most discriminative parts in the feature learning process, therefore, enhancing the information exploitation of the neglected regions. The knowledge guidance module compares the image-based representation generated from the peak suppression module with the learnable knowledge embedding set to obtain the knowledge response coefficients. Afterwards, it formalizes the knowledge learning as a classification problem using response coefficients as the classification scores. Knowledge embeddings and image-based representations are updated during training simultaneously so that the knowledge embedding includes a large number of discriminative clues for different images of the same category.  Finally, we incorporate the acquired knowledge embeddings into the image-based representations as comprehensive representations, leading to significantly higher recognition performance. Extensive evaluations on the six popular datasets demonstrate the advantage of the proposed method in performance. The source code and models will be available online after the acceptance of the paper.
\end{abstract}


\end{frontmatter}

\section{Introduction}


Aiming to distinguish the objects belonging to multiple sub-categories of the same meta-category, fine-grained image recognition has been one of the most fundamental problems in the computer vision and multimedia communities \cite{Han2018Attribute, He2018Only, Li2018I, Zhao2017Diversified, ARAUJO2022427}.
It is essential for a wide range of downstream applications such as rich image captioning \cite{Hendricks2016Deep}, image generation \cite{Bao2017CVAE}, machine teaching \cite{Aodha2018Teaching}, fine-grained image retrieval \cite{Pang2020Solving}, food recognition \cite{Min2019Ingredient, Min2017Being}, and food recommendation \cite{Min2020Food}.

Fine-grained image recognition is challenging due to subtle inter-class differences and significant intra-class variances. Most existing methods only consider the problem of fine-grained recognition from the perspective of obtaining the discriminative characteristics of a single image but ignore the clues provided by multiple images. In order to take the clues of multiple images into consideration, we try to explain the difficulty of fine-grained image recognition with a new perspective and attribute it to fragmented discriminative clues.

\begin{figure}
\centering
 \includegraphics[scale=0.53]{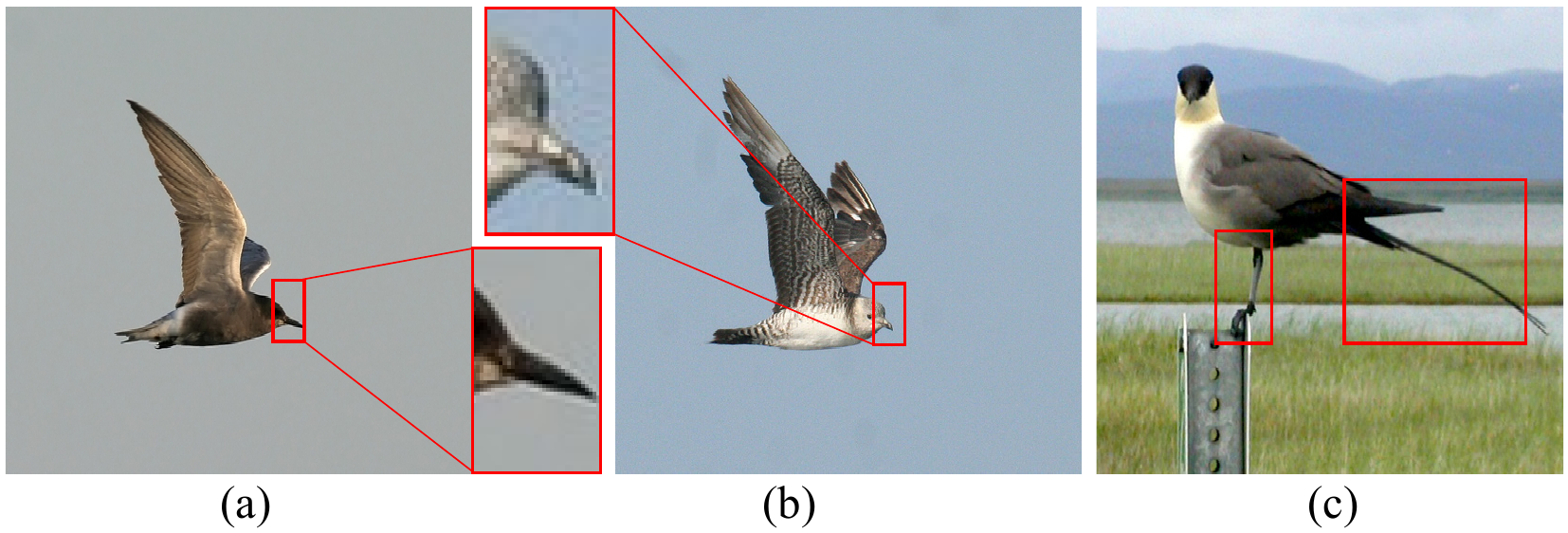}
  \caption{Some fine-grained bird images sampling from the CUB-200-2011 dataset. (a) is Black Tern, both (b) and (c) are Long-tailed Jaeger. Discriminative parts are annotated by red boxes. The details in the red boxes of (a) and (b) are magnified next to the image. Discriminative clues are distributed in multiple regions of the image, and the clues of a single image are usually incomplete.
}
  \label{fig:fragment}
\end{figure}

The fragmentation here has two implications: (1) From the perspective of a single image, discriminative clues appear in different local areas since the inter-class differences could be subtle; As shown in Fig. \ref{fig:fragment} (a) and (b), the long-tailed jaeger is similar to the black tern overall, but the beak of long-tailed jaeger is curved, and the beak of the black tern is straight. This kind of discriminative part is usually tiny and distributed in different image regions, as shown in Figure \ref{fig:fragment} (c). (2) From the perspective of multiple images, each image contains only a part of the discriminative information about the category depending on different poses, scales, and rotations, due to significant intra-class variances.  As shown in Fig. \ref{fig:fragment} (b) and (c), these two pictures are long-tailed jaeger, the beak of the bird in (c) is difficult to distinguish, and there are no bird claws in (b). Therefore, the discriminative information contained in each image is incomplete. 

To find and aggregate fragmented clues is the key to fine-grained image recognition.
Despite their impressive results, existing methods usually consider fine-grained image recognition only from few regions, ignoring many informative details in other regions and other associated images. For instance, if the beak of a specific bird is very different from other birds, the model may pay much attention to the beak of the bird while ignoring the claws and tail of the bird. When this happens, the model could easily make mistakes when the beak of the bird is not visible.

Given this challenge, we propose a Transformer with Peak Suppression and Knowledge Guidance (TPSKG) for fine-grained image recognition. The proposed Peak Suppression (PS) module uses the transformer architecture to integrate the local information and explores a training routine to increase the diversity of discriminative features. This PS module is designed to obtain as many discriminative clues as possible from a single image. Simultaneously the proposed Knowledge Guidance (KG) module incorporates the learnable knowledge embedding into the image-based representation for a comprehensive representation. This KG module is used to aggregate discriminative information from multiple images. 

Specifically, we are inspired by ViT \cite{Dosovitskiy2021Image} and use a transformer architecture to tackle the fine-grained image recognition problems. The input image is reshaped into a patch sequence without overlap and then linearly mapped to the sequential tokens. The transformer encoder uses the self-attention mechanism to integrate the information of the different tokens to obtain a global representation. Instead of integrating all the token information like the original ViT, we deliberately remove the most discriminative token based on the value of the attention weight map in training to penalize strongly discriminative learning and enforce the network to pay attention to other neglected informative areas for keeping the fine-grained representation diversity.
 
After that, we use a knowledge embedding set to explicitly express the discriminative clues of the same category from different images and formalize the learning of knowledge embedding as a classification task. The knowledge guidance module measures the similarity of the knowledge embeddings and image-based representations generated from the peak suppression module to obtain the knowledge response coefficients. We use the knowledge response coefficients as the classification scores directly and use the category label as ground truth to supervise the knowledge learning. The image-based representations become more discriminative through the joint training of fine-grained classification and knowledge learning tasks, and the knowledge embeddings are also concurrently updated through iterations. The learning procedure of knowledge embeddings covers the entire training dataset, therefore these embeddings become the comprehensive representations containing various subtle and slight characteristics of all categories. Finally, we obtain the knowledge-based representations computed from the knowledge embedding set along with the knowledge response coefficients, and inject them into the image-based representations. The proposed knowledge embedding learning and exploitation lead to a significant boost for recognition performance.

\begin{figure}
\centering
 \includegraphics[scale=0.49]{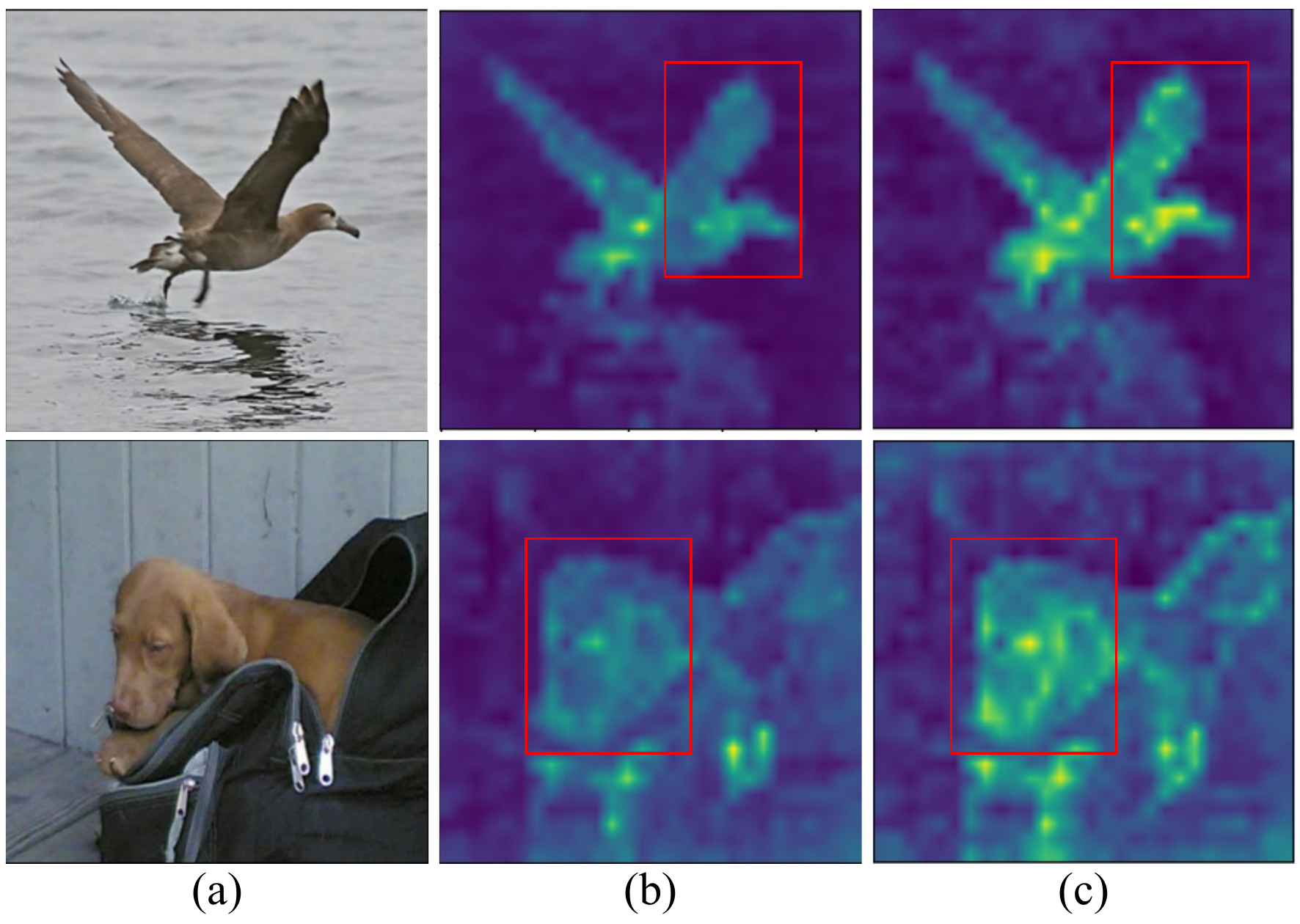}
  \caption{The effects of the proposed approach for some samples from the CUB-200-2011 and Stanford Dogs datasets. (a) is the original fine-grained image, (c) and (b) are the attention weights obtained from the vision transformer with and without the proposed method (TPSKG). The parts of the proposed method that are significantly different from the original method are annotated by red boxes.}
  \label{fig:teaser}
\end{figure}

To verify the effectiveness of our method, we conduct extensive experiments on the six popular benchmarks for the fine-grained image recognition task. Quantitative experimental results demonstrate that the proposed method can achieve competitive performance compared to the state-of-the-art approaches. As shown in Fig. \ref{fig:teaser}, qualitative experimental results demonstrate the advantages of our method in covering more informative areas and increasing the diversity of expression at the same time. The quantitative analysis and visualization of knowledge embedding also illustrate the effectiveness of category-related knowledge embedding learning.

In summary, we make the following main contributions:

(1) We provide a new perspective that the difficulty of fine-grained recognition lies in fragmented discriminative clues. This perspective helps consider not only multiple regions from a single image but also multiple images.

(2) We propose a vision transformer architecture with peak suppression and knowledge guidance for the fine-grained image recognition task. Peak suppression effectively increases the diversity of image representations via aggregating the local features from multiple regions from a single image.  Knowledge guidance optimizes the final representations with the knowledge embeddings learning from multiple images.

(3) We formalize the knowledge learning as a classification problem and directly use the similarity between knowledge embeddings and image-based representations as the classification score to update the knowledge embeddings related to the category.

(4) We conduct extensive quantitative and qualitative experiments to demonstrate the effectiveness of the proposed method, which achieves competitive performance compared to the state-of-the-art approaches on six public datasets.

The rest of this paper is organized as follows. Section \ref{sec.rel} reviews the related works. Section \ref{sec.frame} elaborates on the proposed framework. Experimental results and analysis are reported in Section \ref{sec.exp}. Finally, we conclude the paper in Section \ref{sec.con}.

\section{Related Works} \label{sec.rel}
This section introduces the most related researches into the following categories: the fine-grained image recognition task and the vision transformer architecture.
\subsection{Fine-grained Image Recognition}

There are two prevailing paradigms in the current research in fine-grained image recognition. 
One is the local identification, and the other is the global discrimination.

Local-identification approaches focus on locating the discriminative semantic parts of fine-grained objects to identify the subtle differences among different object categories and construct mid-level representations corresponding to these parts for the final classification.
Early works \cite{Huang2016Part,Donahue2014DeCAF} used strong supervised mechanisms with part bounding box annotations to learn localizing the discriminative parts. However, the part annotation is time-consuming.
Recent researches \cite{Min2019Survey,Ge2019Weakly,Jiang2020Multi, Wang2021} focused on weakly supervised recognition methods with only image-level labels to obtain accurate part localization to solve this problem. 
Some patch-based methods \cite{Yang2018Learning, Liu2021Plant, Liu2020Filtration} first initialize abundant region proposals and select the discriminative parts based on a specific strategy. 
There are also attention-based ways to localize the corresponding high areas related to the image label, such as  \cite{Peng2018Object,Zheng2020Learning, Wang2020Weakly, Min2019Ingredient}.  

Global-discrimination approaches generally learn the embeddings using a specific distance metric so that samples from the same category can be pulled close to each other while samples from different categories are pushed apart.
For example, a bilinear model is used in \cite{Lin2018Bilinear} to learn the interacted feature of two independent CNNs, which achieves remarkable fine-grained recognition performance. However, the exceptionally high dimensionality of bilinear features still makes it impractical for realistic applications.
Chen \textit{et al.} \cite{Chen2019Destruction} enforced the classification network to pay more attention to discriminative regions for spotting the differences by destructing and reconstructing the input image.
Sun \textit{et al.} \cite{Sun2020Fine} masked the most salient features for the input images to force the network to use more subtle clues for its correct classification.
Zhuang \textit{et al.}\cite{Zhuang2020Learning} learned a mutual feature vector to capture semantic differences in the input image pair.

Unlike the methods described above, we consider the discriminative but not the most significant part of a single image, but also emphasize the discriminative information aggregation in different images. Hence, we propose a vision transformer with peak suppression and knowledge guidance, which can effectively increase the richness of fine-grained representations in a local area and effectively aggregate patch features.
Simultaneously, it emphasizes the learning and utilization of the knowledge of distinguishing characteristics between different image samples.

\subsection{Vision Transformer}

The transformer architecture by \cite{Vaswani2017Attention} is proposed to deal with the sequential data in the field of natural language processing \cite{Devlin2019BERT, Brown2020Language}. Inspired by the breakthroughs of transformer architectures in the field of natural language processing, researchers have recently applied transformer to computer vision tasks, such as image recognition \cite{Dosovitskiy2021Image, Touvron2020Training}, object detection \cite{Carion2020End, Zhu2020Deformable}, segmentation \cite{Ye2019Cross}, image super-resolution \cite{yang2020learning}. For example, Cordonnier \textit{et al.} \cite{Cordonnier2020Relationship} proved that a multi-head self-attention layer with a sufficient number of heads is at least as expressive as any convolutional layers. They extracted patches from the input image and applied full self-attention on top. 
iGPT \cite{Chen2020Generative} applies transformers to image pixels after reducing the image resolution and color space.
It is worth noting that the Vision Transformer (ViT) \cite{Dosovitskiy2021Image} is a pure transformer that performs well on the image classification task when applied directly to the sequences of image patches. Based on ViT, Touvron \textit{et al.}\cite{Touvron2020Training} transferred the model to the fine-grained visual categorization and achieved competitive performance. 

To sum up, the vision transformer maps group pixels into a small number of visual tokens, representing a semantic concept in the image. These visual tokens are used directly for image classification, with the transformers being used to model the relationships among tokens. Our work is inspired by ViT and adopts the same method as ViT to build a transformer. Despite the efficiency of iGPT, ViT, and DeiT, these works only fine-tune the model on the fine-grained datasets directly to evaluate the effectiveness of the model for transfer learning and ignore the characteristics of the fine-grained image recognition task. Different from the above methods, we consider the characteristics of the fine-grained image and focus on the specific recognition task.

\section{Framework} \label{sec.frame}

\begin{figure*}
\centering
 \includegraphics[width=\textwidth]{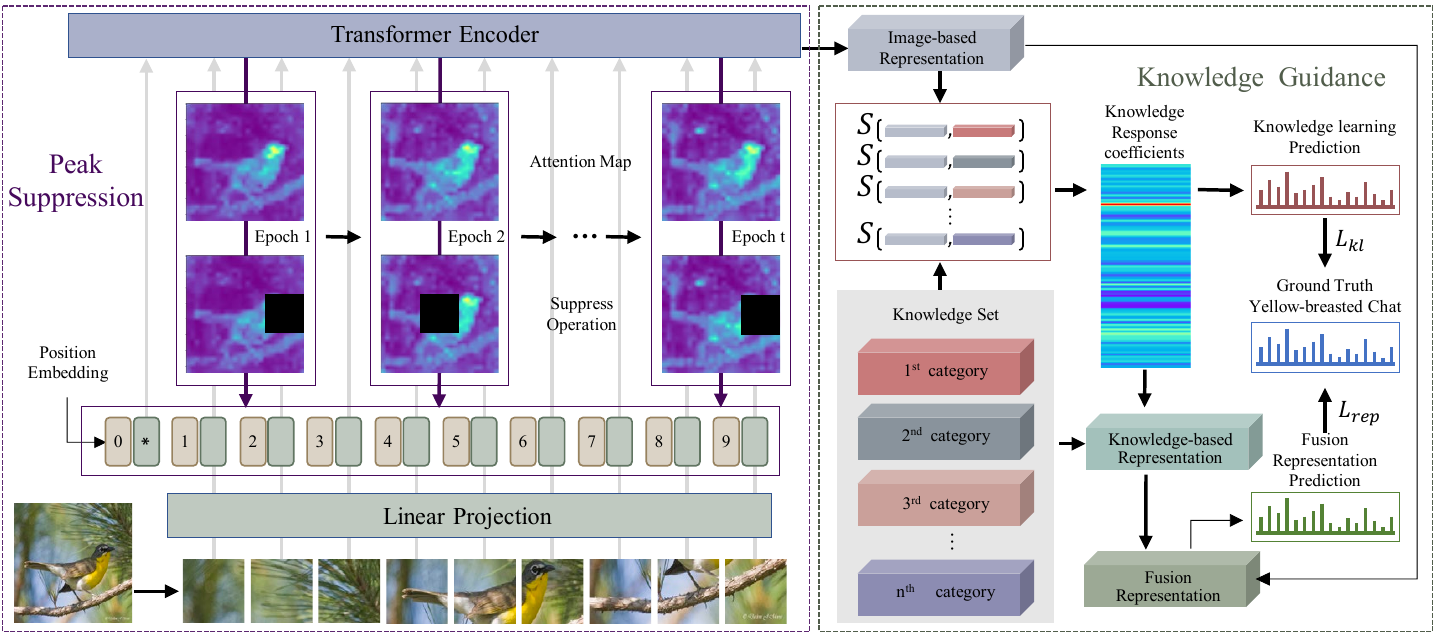}
  \caption{Overview of our framework. Here we visualize the case of peak suppression and knowledge guidance given a training batch with a image and its corresponding label Yellow-breasted Chat. $S(\cdot)$ means the similarity function. Only the presentation label is used to predict in testing.}
  \label{fig:frameworks}
\end{figure*}

This section introduces the proposed framework, which is a transformer architecture for the fine-grained recognition task. 
As shown in Fig. \ref{fig:frameworks}, this framework mainly consists of two components,  namely  Peak Suppression (PS) and Knowledge Guidance (KG).
PS takes images as input and outputs the suppressed image-based representations to the KG module.  The KG module takes the image-based representations and learns the knowledge embeddings, finally uses the fusion representations for the recognition task. 
Section \ref{sec.ps} introduces PS and Section \ref{sec.kg} details KG.

\subsection{Peak Suppression} \label{sec.ps}
Inspired by the effectiveness of the diversification block on convolutional neural networks, we proposed the peak suppression module on the transformer architecture to pay more attention to the other informative parts and obtain more diverse expressions by suppressing the most discriminative regions. Different from the CNNs-based method of directly operating feature maps using the category-specific activation maps, the transformer-based method cannot achieve the goal by directly removing the most significant corresponding token in the last layer because all tokens have interacted during the feedforward process in multi-head attention layers. Therefore, we can only use the attention map to backtrack to the input image space and then mask salient image regions in the image space. 

Formally, we follow the settings of \cite{Dosovitskiy2021Image} and use the ViT as the backbone. Let $x \in \mathbb{R}^{H \times W \times C}$ denotes a given training image where $(H, W)$ is the resolution of the image, $C$ is the number of channels. 
The image $x$ is reshaped into a sequence of flattened 2D patches $x_{p} \in \mathbb{R}^{N \times P^{2}\times C}$, the resolution of each image patch is $(P, P)$, and $N = HW/P^{2}$ is the resulting number of patches. These patches are converted to $D$ dimensions embedding $x_{p}E \in \mathbb{R}^{N \times D}$ as input tokens through a trainable linear projection. Attaching the learnable embedding class token $z_{0}^{0}$, there are a total of $N+1$ tokens. Position embeddings $E_{pos}$ are added to the patch embeddings $E_{pos} \in \mathbb{R}^{(N+1) \times D}$ to retain the positional information. The transformer encoder takes the $z_{0}$ as input and outputs $z_{L}$ and the attention weight $M_{L}$, where $L$ means the transformer encoder is composed of a stack of $L$ identical layers. 
Each layer consists of multi-head self-attention (MSA) and MLP blocks. Layernorm (LN) is applied before every block and residual connections after every block.

\begin{equation}\label{equ.originaltransformer}
\begin{aligned}
z_{0} &= [x_{class};x_{p}^{1}E;x_{p}^{2}E;...;x_{p}^NE] + E_{pos}, \\
z'_{l} &= MSA(LN(z_{l-1})) + z_{l-1}, l = 1...L, \\
z_{l} &= MLP(LN(z'_{l})) + z'_{l}, l = 1...L.\\
\end{aligned}
\end{equation}

We use the Attention Rollout technique \cite{Abnar2020Quantifying} to acquire the attention map from the output token to the input space. Given a transformer with $L$ layers, we need to flow the attention from all positions in the final layer $L$ to all positions in layer 1.
At every transformer layer, we average attention weights at each layer over all heads and get the weight matrix $M_{l}$ that defines the attention value flows from all tokens in the previous layer to all tokens in the $l$ layer.
Considering that there are residual connections in the backbone, we deal with them by adding the identity matrix $I$ to the attention matrices and re-normalize the attention weights to keep the total attention in the range of 0 to 1. Finally, we recursively multiply the weight matrices of all layers. 
\begin{equation}\label{equ.attemap}
\tilde{M}_{l} =\left\{\begin{matrix}
(M_{l} + I)\tilde{M}_{l-1} &  if  l>1, \\  
M_{l}+ I & if  l=1 .
\end{matrix}\right. 
\end{equation}
In this equation, $\tilde{M}_{l}$ is attention rollout of the $l$ layer, $M_{l}$ is raw attention of the $l$ layer, and the multiplication operation is matrix multiplication. The $\tilde{M}_{L}$ illustrates the mixing of attention among tokens across all layers. 

Let $B \in \mathbb{R}^{N+1}$ denote the binary suppressing mask for the input tokens. Each element in mask $B$ is in the domain $\{0, 1\}$, where 0 indicates the corresponding location will be suppressed while 1 means that no suppression will take place. Note that the $B^{0}$ is always 1 because it corresponds to the class token. 

After obtaining the attention map $\tilde{M}_{L}\in \mathbb{R}^{N}$, we compute the $B$ by traversing the entire attention map and finding the position of the largest response.
\begin{equation}\label{equ.findp}
B^{i} =\left\{\begin{matrix}

0 &  if  \tilde{M}_{L}^{i-1} = max(\tilde{M}_{L}) \\  
1 &  otherwise. 
\end{matrix}\right. 
\end{equation}
where  $i \in \{1, 2, ..., N\}$. In order to remove the influence of peaky token, we remove it from the forward process. 

\begin{equation}\label{equ.remove}
\begin{aligned}
\hat{z}_{0} = [x_{class};x_{p}^{1}E;x_{p}^{2}E;...;x_{p}^NE] * B + E_{pos},
\end{aligned}
\end{equation}
where $*$ denotes element-wise multiplication. After the forward process, the transformer encoder outputs $y$ as the image-based representation.
\begin{equation}\label{equ.newtransformer}
\begin{aligned}
\hat{z}_{l}' &= MSA(LN(\hat{z}_{l-1})) + \hat{z}_{l-1}, l = 1...L, \\
\hat{z}_{l} &= MLP(LN(\hat{z}_{l}')) + \hat{z}_{l}', l = 1...L.\\
y &= LN(\hat{z}_{L}^{0}),
\end{aligned}
\end{equation}
where $\hat{z}_{L}^{0}$ denotes the class token vector of the output of $L$ layer transformer encoder after peak suppression.
We implement the remove of Equation (\ref{equ.remove}) by setting $-\infty$ to the suppressed token vectors.
After the softmax layer of Equation (\ref{equ.newtransformer}), the responses of these tokens will be close to 0. 

To sum up, we suppress the peaky token based on the attention maps in the training phase.  By suppressing the tokens, the network is forced to find the other informative regions instead of the most discriminative regions in the image. The increase of the feature diversity can improve the performance of the network in the test phase.

\subsection{Knowledge Guidance} \label{sec.kg}

After obtaining the diversified discriminative clues of a single image, the knowledge guidance module fuses the information of multiple images to get a more comprehensive feature representation. The knowledge guidance module first learns the knowledge embeddings related to the category. To this end, we propose a novel learning method that abstracts the learning of knowledge embeddings as a classification problem and directly uses the similarity between knowledge embeddings and image-based representations as the classification score. Subsequently, the knowledge guidance module injects the obtained knowledge embeddings into the image-based representations to get a more comprehensive expression. Therefore, the knowledge guidance module contains two tasks, one is knowledge learning, and the other is knowledge exploiting. We first introduce the knowledge learning task.

\subsubsection{Knowledge Learning} 
To enforce the networks to learn the knowledge embedding for each category, we treat the knowledge learning task as a classification task. 
Considering the multi-class fine-grained image recognition task, let $\mathcal{F}=\{f^{g}\}_{g=1}^{G}$ denotes the fine-grained label set containing all $G$ fine-grained labels and  $\mathcal{X}=\{x^j\}_{j=1}^{J}$ denotes the image training dataset containing the total $J$ images. Through the peak suppression module of transformer encoder, we can acquire the representation set $\mathcal{Y}=\{y^j\}_{j=1}^{J},y^j \in \mathbb{R}^D$ from the training dataset.
Randomly initializing a knowledge embedding set of the $D$-dimension knowledge embedding $\mathcal{K} =\{k^{g}\}^G_{g=1}, k^{g} \in \mathbb{R}^D$, each $k^{g}$ means the knowledge embedding of the category $f^{g}$.

Given a image $x^{j}$ with the corresponding ground-truth fine-grained label $f^{g}$, the transformer encoder outputs its representation $y^j \in \mathbb{R}^D$. 
The knowledge embedding set tries to distinguish which category this representation $y^j$ belongs to by judging the similarity between this representation $y^j$ and every knowledge embedding $k^{g}$ in the $\mathcal{K}$. We obtain a knowledge response coefficients $r^j \in \mathbb{R}^G$.
\begin{equation}\label{equ.response}
r^j= Softmax(S(y^j,\mathcal{K})),
\end{equation}
where $S(\cdot)$ is a similarity function. We have tried a variety of methods to calculate similarity, such as the neural networks and element-wise multiplication, and the results are not significantly different. In addition, the computational complexity of element-wise multiplication is much smaller. Therefore, without loss of generality, the element-wise multiplication is adopted in our experiments.

We then directly convert the knowledge response coefficients $r^j$ into one-hot for a supervised learning. We define the knowledge learning loss as
\begin{equation}\label{equ.lossatte}
Loss_{kl} = CrossEntropy(Onehot(r^j), f^g ).
\end{equation}

We use $Loss_{kl}$ to supervise knowledge learning processing to update the knowledge embedding set.
The knowledge embeddings have gradually become the common ground of different instances of the same category during the training procedure. As the representations become more discriminative in training, the knowledge embeddings become better to express the category. Furthermore, in this training procedure, the knowledge learning task considers the representations of all training images, making the knowledge embeddings more comprehensive in the expression of corresponding categories.

\subsubsection{Knowledge Exploiting}
In order to effectively use knowledge, we first obtain a knowledge-based representation based on the knowledge response coefficients and the knowledge embeddings.
The knowledge-based representation $\delta^{j}$ is computed as 
\begin{equation}\label{equ.atteValue}
\delta^{j} = \sum_{g \in G}r^jk^{g}.
\end{equation}

This knowledge-based representation includes a summary of the fine-grained features of different instances in the same category and a summary of the differences in different categories.
 
We use the knowledge to guide the classification by injecting the knowledge-based representation  into the final fusion fine-grained representation:
\begin{equation}\label{equ.featFusion}
u^{j} = FC(LN(y^{j}+\delta^{j})),
\end{equation}
where FC is the fully connective layer.

We define a representation learning loss to supervise the training to obtain the fusion representation: 
\begin{equation}\label{equ.lossrep}
Loss_{rep} = CrossEntropy(u^j, f^g ).
\end{equation}

We optimize the knowledge learning task and the knowledge exploiting task simultaneously so that the image representation and the knowledge can be updated iteratively and promote each other during the training process.

Thus, the total loss function of the whole network can be defined as
\begin{equation}\label{equ.atteLoss}
Loss = Loss_{kl}+ \mu \times Loss_{rep},
\end{equation} 
where $\mu$ is a hyperparameter used to adjust the different emphasis of the two tasks.  

During the training, our method explicitly obtains the knowledge embeddings of the different fine-grained categories. These knowledge embeddings are distinguishable and comprehensive, so we insert them into the image-based representations to increase the comprehensiveness of features for the fine-grained recognition task.

In summary, the peak suppression module aims to consider more regions in a single image to obtain more diverse expressions. Based on the peak suppression module, the knowledge guidance module aims to extract and exploit the category-related embeddings based on the expression of multiple images. Therefore, these two modules are complementary in aggregating fragmented information at different levels. 
In the next Section \ref{sec.exp}, we conduct sufficient experiments to prove the effectiveness of the proposed method. 

\section{Experiments}\label{sec.exp}
\subsection{Experimental Setup}
\subsubsection{\textbf{Datasets}}
We conduct our experiments on six fine-grained image recognition datasets,
including two publicly available bird datasets CUB-200-2011 \cite{Wah2011Caltech} and NABirds \cite{Horn2015Building}, one flower dataset Oxford 102 Flowers \cite{Nilsback2008Automated}, one dog dataset Stanford Dogs \cite{Khosla2011Novel}, and two food datasets ISIA Food-200 \cite{Min2019Ingredient} and ISIA Food-500 \cite{Min-ISIA-500-MM2020}. The detailed statistics about these six datasets including class numbers and train/test distributions are summarized in Table \ref{tab:dataset}.

\begin{table}[h]
\centering
  \caption{Fine-grained image dataset statistics.}
  \label{tab:dataset}
\resizebox{\linewidth}{!}{
\begin{tabular}{@{}|c|c|r|r|@{}}
\hline
Dataset       & \# Class & \# Training & \# Testing \\ \hline
CUB-200-2011  \cite{Wah2011Caltech}  & 200   & 5,994    & 5,794   \\
Oxford Flowers \cite{Nilsback2008Automated} & 102   & 1,020    & 6,149   \\
Stanford Dogs \cite{Khosla2011Novel} & 120   & 12,000   & 8,580   \\
NABirds  \cite{Horn2015Building}     & 555   & 23,929    & 24,633   \\
ISIA Food-200	\cite{Min2019Ingredient}	  &	200   & 118,210   & 59,287    \\
ISIA Food-500	\cite{Min-ISIA-500-MM2020}	  &	500   & 239,379   & 120,143    \\\hline
\end{tabular}}
\end{table}

\subsubsection{\textbf{Implementation Details and Comparison Methods}}
Our method is implemented on the Pytorch platform with four Nvidia V100 GPUs.
The input image size is 448 $\times$ 448 as most state-of-the-art fine-grained image recognition approaches. By following the settings of NTS-NET \cite{Yang2018Learning}, we use data augmentations, including random cropping and horizontal flipping during the training procedure. Only the center cropping is involved in inference. The model is trained with the stochastic gradient descent (SGD) with a batch size of 8 and momentum of 0.9 for all datasets. The learning rate is set to 3e-2 initially, and the schedule applies a cosine decay function to an optimizer step.
In all our experiments, we use ViT-B-16 pre-trained on ImageNet21k as the backbone. For all experiments, we adopt the top-1 accuracy as the evaluation metric.
To demonstrate the advantages of our model, we list the some methods for comparisons.

\begin{itemize}

\item {DVAN  \cite{Zhao2017Diversified}}: Diversified visual attention network which  relieves the dependency on supervised information.
\item {MaxEnt  \cite{Dubey2018Maximum}}: Maximum entropy approach which provides a training routine to maximize the entropy of the output probability distributions.
\item {NTS-NET \cite{Yang2018Learning}}: Navigator-teacher-scrutinizer network which finds consistent informative regions through multi-agent cooperation.
\item {Cross-X  \cite{Luo2019Cross}}:  Multi-scale feature learning with exploiting the relationships between different images and layers.

\item {GHNS \cite{Kim2021}}: A framework that generates features of hard negative samples.
\item {CSC-Net  \cite{Wang2020Category}}:  Category-specific semantic coherency network which semantically aligns the discriminative regions of the same subcategory.
\item {CDL  \cite{Wang2019Weakly}}:  Correlation-guided discriminative learning model which mines and exploits the discriminative potentials of correlations.

\item {MLA-CNN \cite{Ji2021}}: Multi-level attention model which uses the neural activations to generate multi-scale regions which are helpful for the fine-grained categorization.

\item{BiM-PMA \cite{Song2020Bi}}: Progressive mask attention model by leveraging both visual and language modalities.
\item {CIN      \cite{Gao2020Channel}}: Channel interaction network models channel-wise interplay  within an image and across images.
\item {DB      \cite{Sun2020Fine}}: End-to-end network with a diversification block to use more subtle clues.

\item {FDL     \cite{Liu2020Filtration}}: Filtration and distillation learning model with region proposing and feature learning.
\item {PMG     \cite{Du2020Fine}}: Progressive multi-granularity method exploiting information based on the smaller granularity information found at the last step and the previous stage.
\item {API-NET  \cite{Zhuang2020Learning}}: Pairwise interaction network which can progressively recognize a pair of images by interaction.

\item {CPM   \cite{Ge2019Weakly}}: Complementary part model in a weakly supervised manner to retrieve information suppressed by dominant object parts detected by CNNs.

\item {GaRD \cite{Zhao_2021_CVPR}}: Graph-based relation discovery approach to grasp the stronger contextual details.

\item {SnapMix \cite{Huang2021SnapMix}}: Semantically proportional mixing which exploits CAM to lessen the label noise in augmenting fine-grained data.

\item {HGNet    \cite{CHEN2021}}:  Hierarchical gate network  to exploit the interconnection among hierarchical categories.


\item {SCAPNet \cite{Liu2021Learning}}: Scale-consistent attention part network to guide part selection across multi-scales and keep the selection scale consistent.

\item {CTF-CapsNet \cite{Lin2021}}: Coarse-to-fine capsule network to shape an increasingly specialized description. 
\item {MSEC \cite{Zhang2021}}: Multi-Scale Erasure and Confusion which realizes confusion at different scales in images and sub-regions.

\end{itemize}

\subsection{Ablation Analysis}
In this section, we conduct a series of ablation studies on the CUB-200-2011, Stanford Dog, Oxford 102 Flowers, NABirds, ISIA Food-200, and ISIA Food-500 datasets to better understand the designation of the proposed TPSKG. We use the performance of the original ViT-B-16 as the ablation baseline.

\subsubsection{\textbf{Impact of different components}}
To investigate the contribution of each component in the proposed method, we omit different components of TPSKG and report the corresponding top-1 recognition accuracy. From the results reported in Table \ref{tab:ablation}, we can draw the following conclusions: 

\begin{table*}[h]
\centering
\caption{Ablation results of the proposed TPSKG on the fine-grained image datasets.}
\label{tab:ablation}
\resizebox{\textwidth}{!}{
\begin{tabular}{|c|c|c|c|c|c|c|}
\hline
Model   & CUB-200-2011 & Stanford Dog & Oxford Flowers & NABirds & ISIA Food-200 & ISIA Food-500 \\ \hline
ViT-B-16               & 90.4           & 91.4         & 99.2   & 89.6                         & 67.4                               & 59.9                               \\ \hline
w/o Peak Suppression   & 91.0           & 91.8         & 99.3   & 89.9                         & 69.3                               & 62.0                                    \\
w/o Knowledge Guidance & 90.9           & 91.8         & 99.3   & 89.8                         & 68.3                               & 61.2                                    \\ \hline
All TPSKG              & 91.3           & 92.5         & 99.5   & 90.1                         & 69.5                               & 65.4                               \\ \hline
\end{tabular}}
\end{table*}

(1) The recognition accuracy on the CUB-200-2011 dataset drops from 91.3\% to 91.0\% and 90.9\% when omitting the PS module and the KG module respectively, which demonstrates the effectiveness of both of the components for the fine-grained image recognition task. The experimental results on the other five datasets also have similar trends to the results of the CUB-200-2011 dataset, indicating that both the PS module and the KG module can effectively improve the recognition performance. 

(2) The network with only the PS module improves the recognition accuracy of baseline by 0.5\% (90.4\% vs. 90.9\%) on CUB-200-2011 dataset, shows that details other than the most significant part are helpful for the fine-grained image recognition task. At the same time, the network with the KG module improves 0.6\% (90.4\% vs. 91.0\%) on the CUB-200-2011 dataset, showing that integrating the discriminative information of multiple images can effectively improve the recognition performance of the network. This result is consistent with our analysis of discriminative information fragmentation in fine-grained image recognition tasks.

(3) The PS module improves the recognition accuracy of baseline by 1.3\% (59.9\% vs. 61.2\%) and the KG module improves the recognition accuracy of baseline by 2.1\% (59.9\% vs. 62.0\%) on the ISIA Food-500 dataset, the combination of these two modules can improve the recognition accuracy of baseline by 5.5\% (59.9\% vs. 65.4\%). This experimental phenomenon shows that the PS module and the KG module are complementary and can promote each other. The more diverse the expression obtained by the peak suppression module, the stronger the expression ability of embedding learned by the knowledge guidance module.

\begin{figure}[h]
\centering
 \includegraphics[scale=0.48]{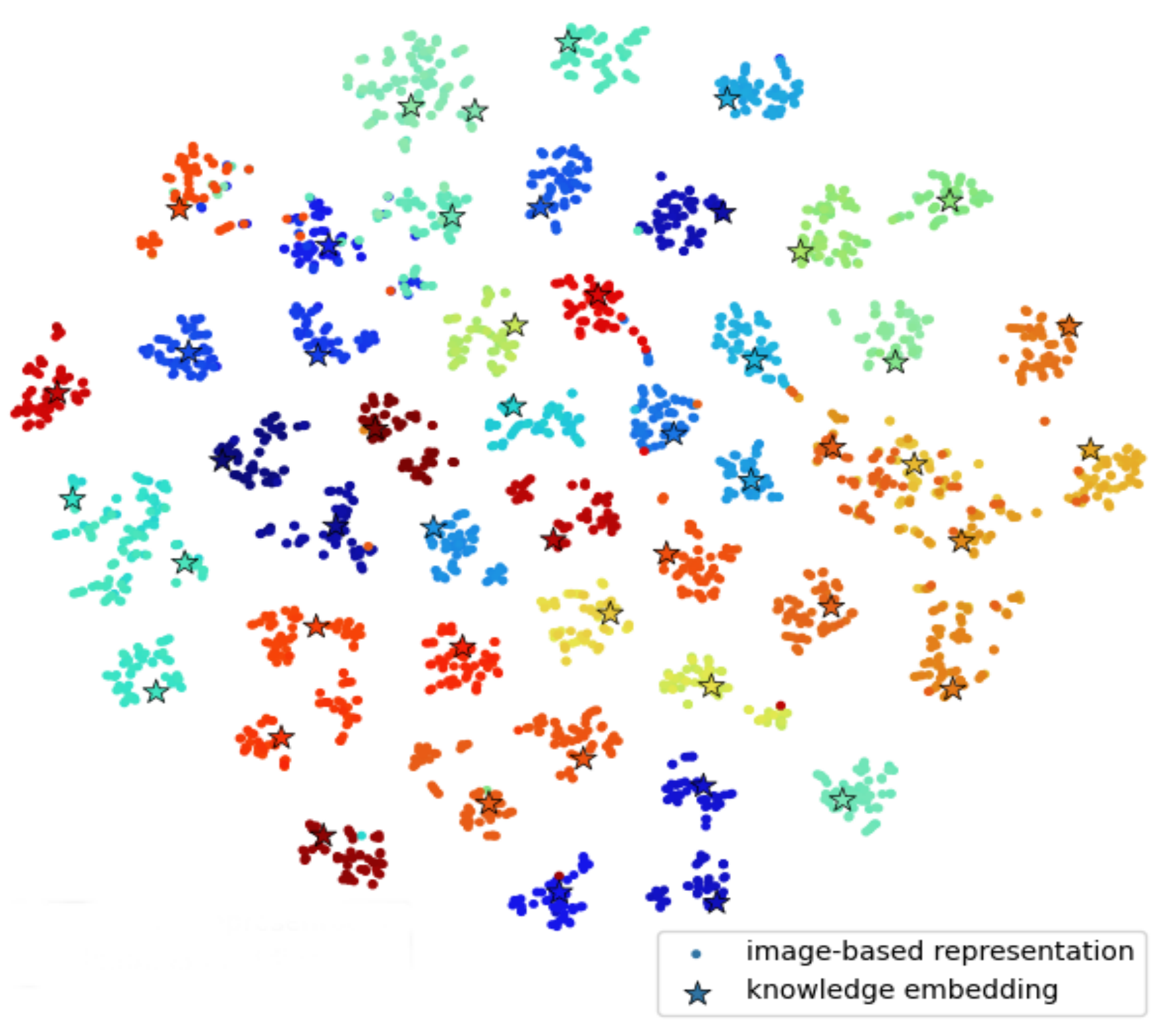}
  \caption{The t-SNE visualizations of image representations and knowledge embedding of 50 sample categories from the CUB-200-2011 dataset. Each dot stands for an image representation, and each star represents the knowledge embedding of a category. The color indicates the categories.}
  \label{fig:knowledge}
\end{figure}

\subsubsection{\textbf{Visualizations of knowledge embedding}}
The knowledge embedding is a category-related feature representation learning with the similarity as the classification score directly. The embedding becomes the most similar to the feature expression in the corresponding category and the most dissimilar to the feature expression in other categories. As shown in Fig. \ref{fig:knowledge}: (1) The knowledge-based representation and the image-based representation are in the same subspace, which is convenient for feature fusion. (2) These two representations of the same category are close in the subspace, indicating that knowledge-based representation can express category-related information. (3) The knowledge-based representation is separable in the subspace, so it is helpful for recognition tasks.

\subsubsection{\textbf{Choice of $\mu$:}}
\begin{figure}[h]
\centering
 \includegraphics[scale=0.8]{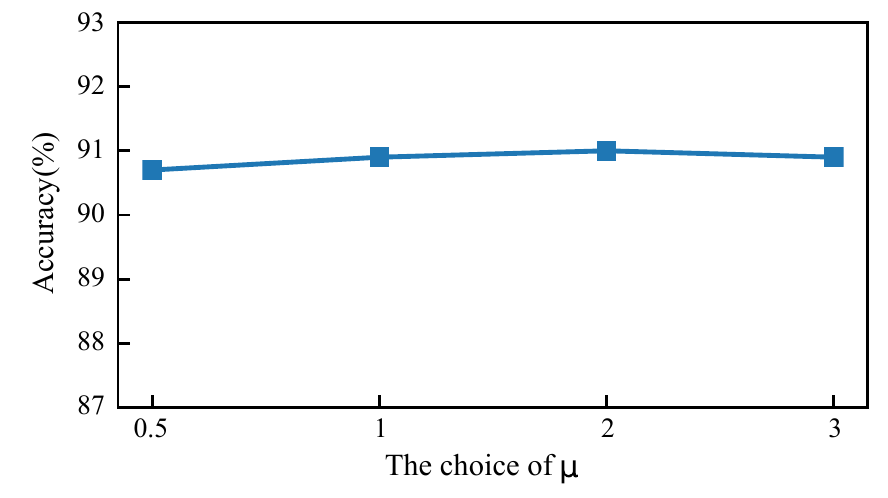}
  \caption{Performance on the choice of $\mu$ on the CUB-200-2011 dataset.}
  \label{fig:choiceu}
\end{figure}

Since Equation \ref{equ.atteLoss} requires selecting a hyperparameter $\mu$, it is essential to study the influence of classification performance on the choice of $\mu$. We conduct this experiment for four different $\mu$ on the CUB-200-2011 dataset. As shown in Fig. \ref{fig:choiceu}, the experimental results show that (1) The performance is relatively robust to the choice of $\mu$ generally. (2) The model performs better when the weight of $Loss_{rep}$ is slightly larger than $Loss_{kl}$. The probable reason is that the learning of knowledge relies on the combined effect of representation and label. A slightly larger weight of $Loss_{rep}$  allows the network to learn the discriminative feature representation preferentially. Because the performance of the model is not sensitive to the choice of $\mu$, the weighting coefficient in Equation \ref{equ.atteLoss} is empirically chosen to be $\mu$=2 in all the following experiments.
 
\subsubsection{\textbf{Computation complexity}}
Since the computation and memory cost can be heavy for global attention in vision transformer architecture, we compare the proposed modules to the original ViT, both in terms of the number of parameters and MACs (Multiply–Accumulate Operations). As shown in Table \ref{tab:macs}, the proposed KG module achieves better performance compared to the original ViT without significantly increasing computational complexity. Although the proposed PS module increases the computational complexity in the training phase, it does not increase the computational complexity at all in the inference phase, and at the same time improves the recognition performance. The number of parameters has not increased significantly.
\begin{table}[h]

\caption{Comparison results on Cub-200-2011 dataset.}
\label{tab:macs}
\resizebox{\linewidth}{!}{
\begin{tabular}{|c|c|c|c|c|c|}
\hline
Models    & \begin{tabular}[c]{@{}c@{}}Input\\ Size\end{tabular} & \begin{tabular}[c]{@{}c@{}}Params\\ (M)\end{tabular} & \begin{tabular}[c]{@{}c@{}}Training\\ MACs(G)\end{tabular} & \multicolumn{1}{l|}{\begin{tabular}[c]{@{}l@{}}Inference\\ MACs(G)\end{tabular}} & \begin{tabular}[c]{@{}c@{}}Accuracy\\ (\%)\end{tabular} \\ \hline
ViT-B-16  & 448$\times$448    & 86.4                                                 & 67.14                                                      & 67.14                                                                            & 90.4                                                    \\
ViT w/ PS & 448$\times$448    & 86.4                                                 & 134.4                                                      & 67.14                                                                            & 90.9                                                    \\
ViT w/ KG & 448$\times$448    & 86.6                                                 & 67.14                                                      & 67.14                                                                            & 91.0                                                    \\ \hline
TPSKG     & 448$\times$448    & 86.6                                                 & 134.4                                                      & 67.14                                                                            & 91.3                                                    \\ \hline
\end{tabular}}
\end{table}

\subsection{Comparison with State-of-the-art}
For further verification for the TPSKG, we compare our method to the state-of-the-art methods on the six publicly available fine-grained datasets in this section.

\subsubsection{\textbf{CUB-200-2011}}
We compare the proposed method against many state-of-the-art fine-grained recognition models on CUB-200-2011, as shown in Table \ref{tab:cub}. The results show the following conclusions.

(1) Overall, the proposed TPSKG performs better than the state-of-the-art fine-grained methods, including the global-discrimination approaches methods MaxEnt \cite{Dubey2018Maximum} and DB \cite{Sun2020Fine}, and the part-based methods NTS-NET \cite{Yang2018Learning} and CPM \cite{Ge2019Weakly}. 

(2) Images with higher resolutions usually contain richer information and subtle details that are important for the fine-grained image recognition task.  According to the literature \cite{Cui2018Large}, higher resolution input images will produce better performance generally. Our method uses a smaller resolution than PMG \cite{Du2020Fine}, API-NET \cite{Zhuang2020Learning}, and CPM \cite{Ge2019Weakly} but achieves a better performance. At the same time, our method is possible to perform better with higher resolution. 

(3) CPM \cite{Ge2019Weakly} has good performance using the stacked BiLSTMs to integrate the patch features. The performance of the original ViT method is equivalent to the CPM, which proves the effectiveness of the transformer in feature aggregation and its potential in the fine-grained recognition task. 

(4) Although the original ViT model has satisfactory performance, we have improved it by 0.9\%. 

(5) Both CIN \cite{Gao2020Channel} and API-NET \cite{Zhuang2020Learning} have achieved good results based on the contrastive learning mechanism. It is worth noting that the improvement from our framework is orthogonal to those works, so the proposed TPSKG  can also benefit from these methods.

\begin{table}[h]
\centering
\caption{Comparison results on CUB-200-2011 dataset.}
\label{tab:cub}
\resizebox{\linewidth}{!}{
\begin{tabular}{|c|c|c|c|}
\hline
Method    & Backbone     & Resolution & Accuracy(\%) \\\hline
MaxEnt  \cite{Dubey2018Maximum}  & DenseNet-161 & -          & 84.9         \\
MLA-CNN \cite{Ji2021}      &  VGG-19       & 448 $\times$ 448    & 85.7         \\
DVAN  \cite{Zhao2017Diversified}  & VGG-16    & 224 $\times$ 224    & 87.1         \\
NTS-NET \cite{Yang2018Learning}  & ResNet-50    & 448 $\times$ 448    & 87.5         \\
Cross-X  \cite{Luo2019Cross}  & ResNet-50    & 448 $\times$ 448    & 87.7         \\
HGNet    \cite{CHEN2021}  & ResNet-50    & 448 $\times$ 448    & 87.9         \\
CIN      \cite{Gao2020Channel} & ResNet-101   & 448 $\times$ 448    & 88.1         \\
MSEC \cite{Zhang2021} & ResNet-50   & 448 $\times$ 448    & 88.3         \\
CDL  \cite{Wang2019Weakly} & ResNet-50   & 448 $\times$ 448    & 88.4         \\
DB      \cite{Sun2020Fine}  & ResNet-50    & 448 $\times$ 448    & 88.6         \\

HGNet    \cite{CHEN2021}  & ResNet-50    & 448 $\times$ 448    & 88.7         \\
CTF-CapsNet \cite{Lin2021}  & ResNet-50    & 448 $\times$ 448    & 88.9         \\
GHNS \cite{Kim2021}         & ResNet-50    & 448 $\times$ 448    & 89.1       \\
FDL     \cite{Liu2020Filtration}  & DenseNet-161 & 448 $\times$ 448    & 89.1         \\
CSC-Net  \cite{Wang2020Category}  & ResNet-50 & 224 $\times$ 224    & 89.2         \\
SCAPNet \cite{Liu2021Learning} & ResNet-50 & 224 $\times$ 224    & 89.5         \\
PMG     \cite{Du2020Fine}  & ResNet-50    & 550 $\times$ 550    & 89.6         \\
GaRD \cite{Zhao_2021_CVPR}  & ResNet-50    & 448 $\times$ 448    & 89.6         \\
SnapMix \cite{Huang2021SnapMix} & ResNet-101    & 448 $\times$ 448    & 89.6         \\
CTF-CapsNet \cite{Lin2021}  & ResNet-50    & 448 $\times$ 448    & 89.7         \\
API-NET  \cite{Zhuang2020Learning} & DenseNet-161 & 512 $\times$ 512    & 90.0         \\
CPM   \cite{Ge2019Weakly} & GoogLeNet    & over 800   & 90.4         \\\hline
ViT-ResNet-50                             & ViT\&ResNet-50     & 448 $\times$ 448    & 89.2         \\
ViT   \cite{Dosovitskiy2021Image}    & ViT-B-16     & 448 $\times$ 448    & 90.4         \\
TPSKG      & ViT-B-16     & 448 $\times$ 448    & \textbf{91.3}         \\\hline
\end{tabular}}
\end{table}

\subsubsection{\textbf{Stanford Dog}}
As can be seen from Table \ref{tab:dog}, our method shows more performance improvement on the Stanford Dog dataset, which is 2.2\% higher than the current state-of-the-art method API-NET \cite{Zhuang2020Learning} without using the contrastive learning mechanism and the high-resolution input. It is worth noting that the performance of the original ViT also exceeds  API-NET by 1.1\%, which reflects that the transformer architecture can be well migrated to fine-grained recognition task.

\begin{table}[h]
\centering
\caption{Comparison results on Stanford Dog dataset.}
\label{tab:dog}
\resizebox{\linewidth}{!}{
\begin{tabular}{|c|c|c|c|}
\hline
Method  & Backbone     & Resolution & Accuracy(\%) \\\hline
DVAN  \cite{Zhao2017Diversified}  & VGG-16    & 224 $\times$ 224    & 81.5         \\
MaxEnt \cite{Dubey2018Maximum} & DenseNet-161 & -          & 83.6         \\
PC-CNN \cite{Dubey2018Pairwise} & DenseNet-161 & 224 $\times$ 224          & 83.8         \\
FDL  \cite{Liu2020Filtration}   & DenseNet-161 & 448 $\times$ 448    & 84.9         \\
MSEC \cite{Zhang2021} & ResNet-50   & 448 $\times$ 448    & 85.6         \\
MLA-CNN \cite{Ji2021}      &  VGG-19       & 448 $\times$ 448    & 86.8         \\
DB   \cite{Sun2020Fine}    & ResNet-50    & 448 $\times$ 448    & 87.7         \\
Cross-X \cite{Luo2019Cross}& ResNet-50    & 448 $\times$ 448    & 88.9         \\
API-NET \cite{Zhuang2020Learning} & DenseNet-161 & 512 $\times$ 512    & 90.3         \\\hline
ViT-ResNet-50                             & ViT\&ResNet-50     & 448 $\times$ 448    & 87.7         \\
ViT  \cite{Dosovitskiy2021Image}   & ViT-B-16     & 448 $\times$ 448    & 91.4         \\
TPSKG    & ViT-B-16     & 448 $\times$ 448    &\textbf{92.5}       \\\hline
\end{tabular}}
\end{table}

\subsubsection{\textbf{Oxford 102 Flowers}}
Unlike BiM-PMA \cite{Song2020Bi}, which uses all 2040 images in the training set and validation set for training, we follow the settings of PC-CNN \cite{Dubey2018Pairwise} and PBC \cite{Huang2017PBC} and only use 1020 images in training set for training to ensure a relatively fair comparison.  As can be seen from Table \ref{tab:flower}, although using fewer images, our method still achieves a 2.1\% performance improvement compared to BiM-PMA. At the same time, our method still improves the recognition performance when the recognition performance of ViT is excellent. 

\begin{table}[h]
\centering
\caption{Comparison results on Oxford 102 Flowers dataset.}
\label{tab:flower}
\resizebox{\linewidth}{!}{
\begin{tabular}{|c|c|c|c|}
\hline
Method  & Backbone     & Resolution & Accuracy(\%) \\\hline
PC-CNN \cite{Dubey2018Pairwise} & DenseNet-161 & 224 $\times$ 224          &  93.6        \\
PBC \cite{Huang2017PBC} & GoogleNet & 224 $\times$ 224          &  96.1        \\
BiM-PMA \cite{Song2020Bi} & VGG-16 & 448 $\times$ 448    & 97.4         \\\hline
ViT-ResNet-50                             & ViT\&ResNet-50     & 448 $\times$ 448    & 98.5         \\
ViT  \cite{Dosovitskiy2021Image}   & ViT-B-16     & 448 $\times$ 448    & 99.2         \\
TPSKG    & ViT-B-16     & 448 $\times$ 448    &\textbf{99.5}       \\\hline
\end{tabular}}
\end{table}

\subsubsection{\textbf{NABirds}}
The NABirds dataset is a larger fine-grained dataset than the CUB-200-2011 dataset containing 48,562 North American bird images. Many methods with complex operations are not easy to experiment on a data set of this order of magnitude. If an image generates thousands of proposals, it means that tens of millions of proposals need to be processed. 
Table \ref{tab:nabirds} reports the performance of several methods on the NABirds dataset. 
DSTL \cite{Cui2018Large} uses the transfer learning strategy for the fine-grained image recognition task consisting of more than one dataset. The result of our method trained on a separate dataset exceeds DSTL by 2.2\% in accuracy. 

\begin{table}[h]
\centering
\caption{Comparison results on NABirds dataset.}
\label{tab:nabirds}
\resizebox{\linewidth}{!}{
\begin{tabular}{|c|c|c|c|}
\hline
Method  & Backbone     & Resolution & Accuracy(\%) \\\hline
PC-CNN \cite{Dubey2018Pairwise} & DenseNet-161 & 224 $\times$ 224          & 82.8         \\
MaxEnt \cite{Dubey2018Maximum} & DenseNet-161 & -          & 83.0         \\
Cross-X \cite{Luo2019Cross} & ResNet-50    & 448 $\times$ 448    & 86.4         \\
HGNet    \cite{CHEN2021}  & ResNet-50    & 448 $\times$ 448    & 86.4         \\
DSTL \cite{Cui2018Large}   & Inception-v3 & 560 $\times$ 560          & 87.9         \\
GaRD \cite{Zhao_2021_CVPR}  & ResNet-50    & 448 $\times$ 448    & 88.0         \\\hline
ViT-ResNet-50                             & ViT\&ResNet-50     & 448 $\times$ 448    & 86.7         \\
ViT  \cite{Dosovitskiy2021Image}   & ViT-B-16     & 448 $\times$ 448    &  89.6            \\
TPSKG    & ViT-B-16     & 448 $\times$ 448    &   \textbf{90.1}          \\\hline
\end{tabular}}
\end{table}

\subsubsection{\textbf{ISIA Food-200}}
In order to further verify the effectiveness and explore the scope of application of our method, we explored the food recognition task on the ISIA Food-200 dataset. Unlike other fine-grained objects, many types of food are non-rigid and lack a fixed spatial structure and semantic pattern. Therefore, it is challenging to capture specific semantic information from food images. Our method attempts to increase the diversity of representations to cover more distinguished areas, which is more effective for non-rigid objects. Simultaneously, our method injects the extracted knowledge into the image-based representation so that a more comprehensive understanding of food categories can be used in the recognition task.

Table \ref{tab:food200} reports the performance of several methods on the ISIA Food-200 dataset.
The ViT is also mediocre on this task. IG-CMAN \cite{Min2019Ingredient} is a patch-based method sequentially localizing multiple informative image regions with multi-scale from category level to ingredient-level guidance in a coarse-to-fine manner. Our method achieves the best 69.5\% without using the multi-scale strategy and outperforms the state-of-the-art method IG-CMAN by 2.0\% in accuracy. This result proves that our method has a more significant performance improvement in complex recognition problems.

\begin{table}[h]
\centering
\caption{Comparison results on ISIA Food-200 dataset.}
\label{tab:food200}
\resizebox{\linewidth}{!}{
\begin{tabular}{|c|c|c|c|}
\hline
Method  & Backbone     & Resolution & Accuracy(\%) \\\hline
ResNet-152  & ResNet-152    & 224 $\times$ 224    & 61.1         \\
DenseNet-161  & DenseNet-161    & 224 $\times$ 224    & 62.6         \\
IG-CMAN \cite{Min2019Ingredient}   & DenseNet-161 & 224 $\times$ 224          & 67.5        \\\hline
ViT-ResNet-50                             & ViT\&ResNet-50     & 448 $\times$ 448    & 62.5         \\
ViT  \cite{Dosovitskiy2021Image}   & ViT-B-16     & 448 $\times$ 448    &    67.4           \\
TPSKG    & ViT-B-16     & 448 $\times$ 448    &  \textbf{69.5}           \\\hline
\end{tabular}}
\end{table}

\subsubsection{\textbf{ISIA Food-500}}
The ISIA Food-500 is a more comprehensive food dataset than the ISIA Food-200 with a larger data volume and higher diversity. 
We evaluated the proposed TPSKG against different fine-grained methods in Table \ref{tab:food500}.
The performance of ViT-ResNet-50 and ViT show a pronounced  decline. The possible reason is that the volume and complexity of the ISIA Food-500 dataset are much higher than that of the ISIA Food-200 dataset. This increase in complexity makes the impact of the loss of local region semantics more significant.
It can also be seen that the proposed method exceeds the original ViT significantly, with a gain of 5.5\% in accuracy. When the performance of the original ViT is poor, our method still achieves competitive performance compared to the state-of-the-art method, which proves that our method does not rely heavily on the performance of ViT. 
The proposed method obtains better accuracy than the SGLANet without a complicated multi-scale mechanism and spatial-channel attention. 
We will try to leverage the multi-scale information to improve performance in future work.

\begin{table}[h]
\centering
\caption{Comparison results on ISIA Food-500 dataset.}
\label{tab:food500}
\resizebox{\linewidth}{!}{
\begin{tabular}{|c|c|c|c|}
\hline
Method  & Backbone     & Resolution & Accuracy(\%) \\\hline
ResNet-152 & ResNet-152    & 224 $\times$ 224    & 57.0         \\
WRN-50 \cite{Zagoruyko2016Wide} & WRN-50    & 224 $\times$ 224    & 60.1         \\
WS-DAN \cite{Hu2019See}   & Inception-v3 & 299 $\times$ 299          & 60.7        \\
NAS-NET \cite{Zoph2018Learning} & ResNet-152    & 331 $\times$ 331    & 60.7         \\
NTS-NET \cite{Yang2018Learning} & ResNet-152    & 448 $\times$ 448    & 63.7         \\
SENet-154 & SENet-154    & 224 $\times$ 224    & 63.8         \\
DCL \cite{Chen2019Destruction} & ResNet-152    & 448 $\times$ 448    & 64.1         \\
SGLANet \cite{Min-ISIA-500-MM2020} & SENet-154    & 224 $\times$ 224    & 64.7         \\\hline
ViT-ResNet-50                             & ViT\&ResNet-50     & 448 $\times$ 448    & 52.7        \\
ViT  \cite{Dosovitskiy2021Image}   & ViT-B-16     & 448 $\times$ 448    &  59.9             \\
TPSKG    & ViT-B-16     & 448 $\times$ 448    &   \textbf{65.4}           \\\hline
\end{tabular}}
\end{table}

%

\begin{figure*}[h]
\centering
 \includegraphics[width=\linewidth]{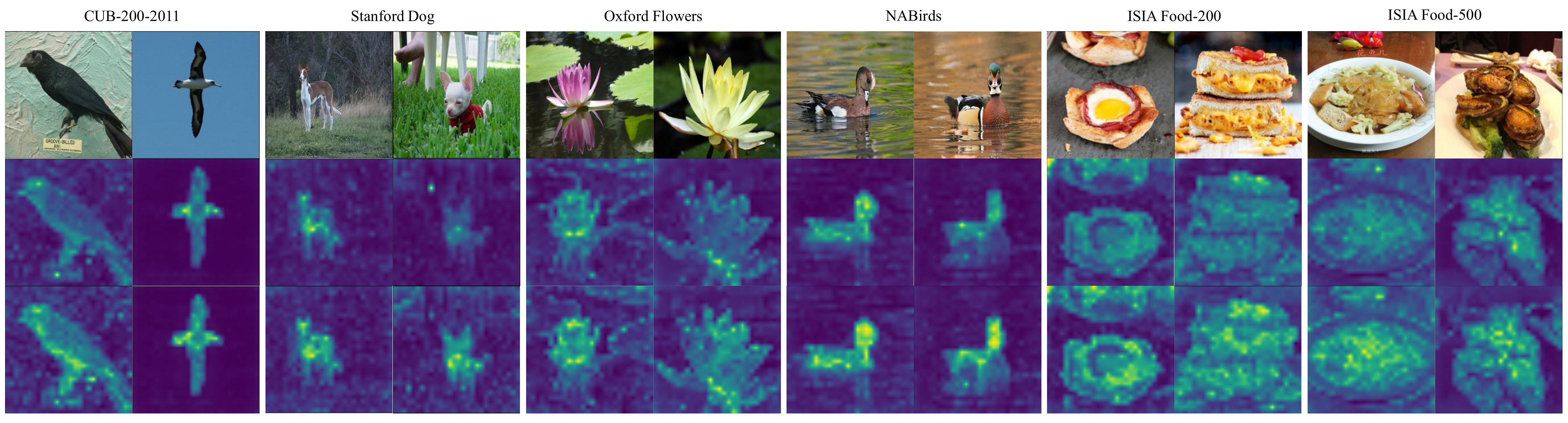}
  \caption{The attention map comparison between our method and the baseline in different datasets. Top to below: original image, attention map of the ViT, attention map of our method. The yellow means high weights and the blue means relatively low weights.}
  \label{fig:visualize}
\end{figure*}

\subsubsection{\textbf{Overall}}
We summarize the results of all datasets to obtain an overall understanding of the proposed method. We can find that 
(1) The recognition results of the different methods for the six different datasets show a very high degree of correspondence, indicating the strong reproducibility. 
(2) The performance of the hybrid model directly generated by the simple combination of ViT and ResNet-50 generally performs poorly on fine-grained image recognition task and even has performance degradation compared to the ViT model. A possible explanation for these results may be the lack of adequate semantic information in small regions. (3) The ViT model is generally suitable for simple fine-grained recognition tasks and obtains close to state-of-the-art results on multiple datasets, but it does not perform well for more complex food recognition tasks. (4) The proposed PS module and KG module have effectively improved the recognition performance, and the proposed method has achieved very superior performance on all datasets. (5) The KG module improves the model performance more significantly than the PS module. The possible explanation for this might be that the benefits of integrating discriminative information in multiple images are more significant than the coverage of more discriminative information areas in one single image.

\subsection{Qualitative Visualization}

\begin{figure*}[]
\centering
 \includegraphics[width=\linewidth]{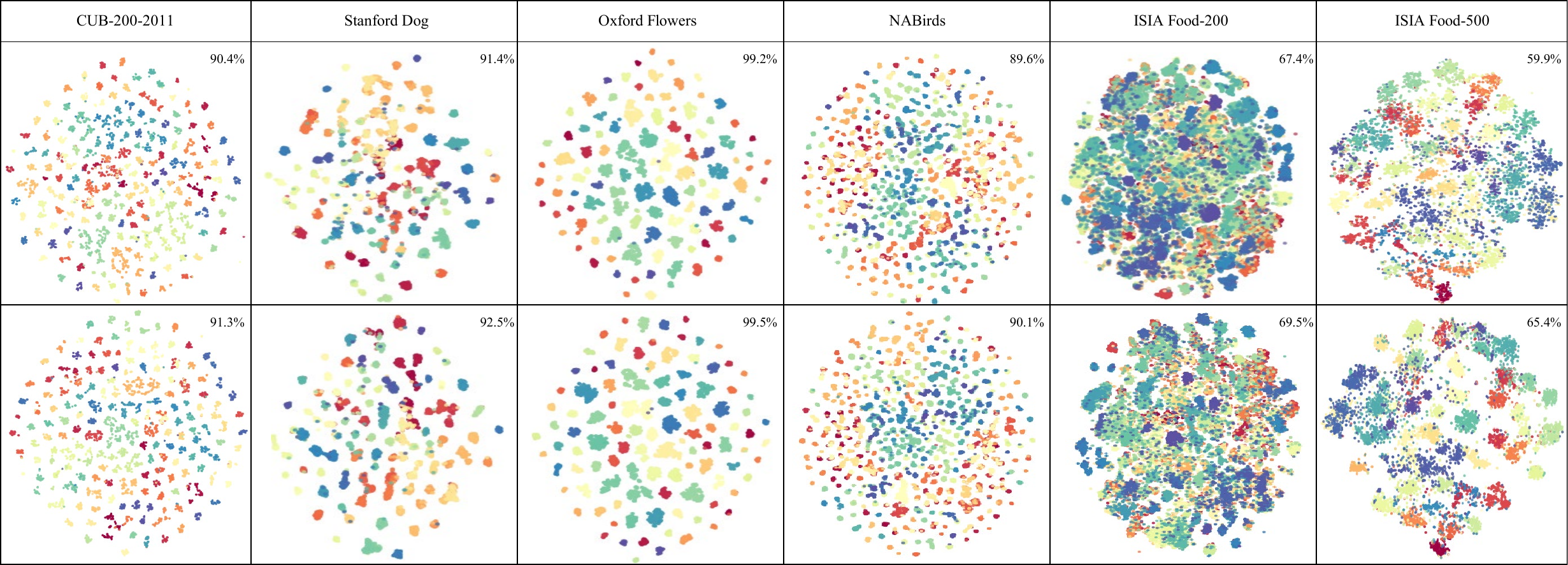}
  \caption{The t-SNE visualization of fine-grained image feature representations of (top row) before injecting the knowledge embedding, (bottom row) after injecting the knowledge embedding on the six fine-grained datasets. Each color represents a different class. The upper right corner shows the accuracy of the corresponding method.}
  \label{fig:tsne}
\end{figure*}

In order to show the effectiveness of the method more intuitively, we visualize the attention maps of the original ViT and TPSKG models for sample images from six different datasets.
As shown in Fig. \ref{fig:visualize}, we find that although the original ViT can also perform localization and recognition, our method is better in both aspects. The attention map of the proposed TPSKG can not only locate the essential parts well but also cover more discriminative areas, which shows that the method is more robust than the original ViT. For the CUB-200-2011, Stanford Dog, Oxford Flowers and NABirds datasets with relatively simple scenes, features without the diversity can complete the recognition task. For relatively complex food recognition tasks on the ISIA Food-200 and ISIA Food-500 datasets, the diversity of features is more important, and our method has improved more obviously, which is consistent with the results of quantitative analysis.

To visually analyze the influence of the knowledge guidance module, we visualize the feature representations before/after injecting the knowledge embedding by employing t-SNE, on the six fine-grained image datasets as shown in Fig. \ref{fig:tsne}. 
The visualized data includes all test set images of CUB-200-2011, Stanford Dogs, Oxford Flowers, NAbirds, ISIA Food-200 datasets and 50 sample categories from the ISIA Food-500 dataset due to excessive data volume. 
Since the classification performance of the first four datasets is very high, the visualized images are not much different. But the latter two more complex food recognition datasets show significant differences. 
The visualization results of food recognition datasets in this figure show obvious intra-class clustering. This proves that the proposed KG module has a strong intra-class aggregation ability. 
Although the t-SNE method cannot prove the method's ability for the recognition task, it can be seen from the local structure that our method has a more vital ability to aggregate features within the class. This investigation confirms that feature representations will get into more separable clusters after injecting the category-related knowledge embedding.

\begin{figure*}[]
\centering
 \includegraphics[width=\linewidth]{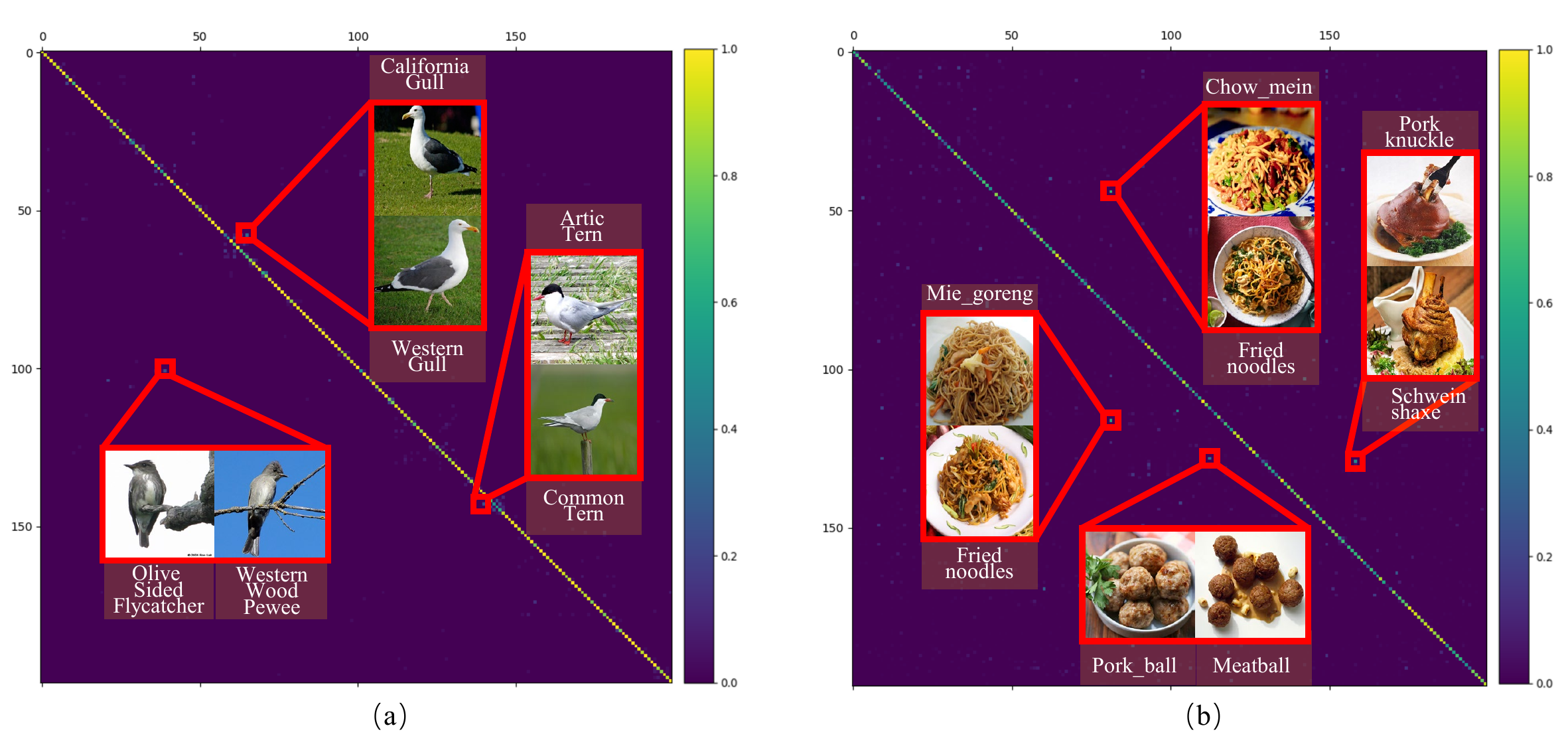}
  \caption{Confusion matrix of our method on the (a) CUB-200-2011 and (b) ISIA Food-200 datasets. Some instances of low recognition rate categories are annotated by red boxes.}
  \label{fig:cf}
\end{figure*}

In addition, we further show the confusion matrix of our method on the CUB-200-2011 and ISIA Food-200 datasets in Fig. \ref{fig:cf}, where the vertical axis shows the ground-truth classes, and the horizontal axis shows the predicted classes. Yellower colors indicate better performance. 
We can see that our method still does not provide perfect performance for some bird and food categories. We enlarge specific regions to highlight the misclassified results and show some samples with low recognition rates. 
As shown in Fig. \ref{fig:cf} (a), some birds of the same meta-category are extremely difficult to distinguish, such as California Gull and Western Gull, Artic Tern and Common Tern. There are also images with similar poses that cannot be recognized well, such as Olive Sided Flycatcher and Wester Wood Pewee, requiring further study and exploration.
From Fig. \ref{fig:cf} (b) we can see that these food categories are very similar in visual appearances, such as Chow mein, Mie Goreng, and Fried noodles. Even humans cannot easily distinguish these food categories based on images. Some food categories have the same ingredients and different cooking techniques, which are difficult to distinguish, such as pork knuckle and the Schweinshaxe.
Many types of food are difficult to classify based on images alone. A possible solution is to combine multiple media formats of information for the recognition task, such as ingredient lists and cooking processes. The transformer architecture has promising applications in multimedia, including the visual field and the NLP field, and provides the possibility for a unified framework.

\section{Conclusion} \label{sec.con}
Fine-grained image recognition is an interesting and fundamental topic. 
In this paper, we investigate the problem of fine-grained image recognition from the perspective of fragmented information integration. Furthermore, we present a transformer with peak suppression and knowledge guidance (TPSKG) for the fine-grained image recognition task. Our method learns the diverse fine-grained representations by the peak suppression module penalizing the most discriminative parts. It then learns the knowledge embedding including a large number of discriminative clues for different images of the same category, and injects them into fine-grained representations leading to significantly higher recognition performance. The proposed network  can be trained end-to-end in one stage, requiring no bounding box/part annotations. Qualitative and quantitative
evaluations on six public fine-grained datasets demonstrate that the proposed TPSKG can achieve competitive performance compared to the state-of-the-art approaches.


\section*{References}

\bibliography{tpskg}

\end{document}